\documentclass[a4paper, 10pt]{ieeeconf}  

\usepackage[english]{babel}
\usepackage[utf8x]{inputenc}
\usepackage[T1]{fontenc}

\usepackage{amsmath}
\usepackage{graphicx}
\usepackage[colorinlistoftodos]{todonotes}
\usepackage[colorlinks=true, allcolors=black]{hyperref}
\usepackage{caption}
\usepackage{subcaption}
\usepackage{multicol}
\usepackage{hyperref}
\usepackage{xfrac}

\title{An Intelligent Cloud Storage Gateway for Medical Imaging}
\author{Carlos Viana-Ferreira, António Guerra, João F. Silva, Sérgio Matos, and Carlos Costa}

\begin{document}
\maketitle

\begin{abstract}
Historically, medical imaging repositories have been supported by indoor infrastructures. However, the amount of diagnostic imaging procedures has continuously increased over the last decades, imposing several challenges associated with the storage volume, data redundancy and availability. Cloud platforms are focused on delivering hardware and software services over the Internet, becoming an appealing solution for repository outsourcing. Although this option may bring financial and technological benefits, it also presents new challenges. In medical imaging scenarios, communication latency is a critical issue that still hinders the adoption of this paradigm. 

This paper proposes an intelligent Cloud storage gateway that optimizes data access times. This is achieved through a new cache architecture that combines static rules and pattern recognition for eviction and prefetching.

The evaluation results, obtained through simulations over a real-world dataset, show that cache hit ratios can reach around 80\%, leading reductions of image retrieval times by over 60\%. 

The combined use of eviction and prefetching policies proposed can significantly reduce communication latency, even when using a small cache in comparison to the total size of the repository. Apart from the performance gains, the proposed system is capable of adjusting to specific workflows of different institutions.

\textbf{Keywords}—Cloud, Medical imaging, Storage gateway, Data access latency, Pattern recognition, Machine learning. 
\end{abstract}

\section{Introduction}

Medical imaging is a very important tool in medical practice, not only for diagnosis but also for patient management and treatment support \cite{beutel2000handbook}. It benefits from technological advances in several areas, including the creation of new imaging modalities and the implementation of the PACS (Picture Archiving and Communication System) concept \cite{huang2011pacs}. PACS refers to systems that are responsible for the acquisition, management, storage, visualization and distribution of medical imaging data \cite{Costa2009273}. Nowadays, these systems proliferate in practically all healthcare institutions, and are also used to support distributed workflows \cite{costa2011dicoogle}.

The Cloud computing paradigm enables on-demand services, such as computing, storage and databases, and answers a current major problem in the medical field: the soaring volume of worldwide healthcare data that results in a Big Data problem \cite{GoliMalekabadi201675}. Much of its interest resides in the fact that, with Cloud computing, computing resources are provided in an elastic way, supporting horizontal scalability, and development and maintenance of cloud software has become easier, more reliable, and safer. Based on this, a tremendous amount of ubiquitous computational power and an unprecedented number of Internet resources and services are used every day as regular commodities. This facility is also being explored for outsourcing of medical imaging services, with two main use cases \cite{silva2012pacs}:
\begin{itemize}
\item PACS archive outsourcing. In-house PACS solutions have high maintenance costs, infrastructure scalability is usually limited and over the years it easily becomes obsolete.
\item Inter-institutional workflows and sharing of medical imaging. For instance, the cloud is excellent for instantiating a teleradiology platform as a service.
\end{itemize}

A major drawback associated with the migration of PACS services to the Cloud is access latency \cite{philbin2011will}, a particularly critical concern in medical imaging scenarios since remote access over the Internet is considerably slower than Intranet connections. Moreover, some studies can amount to a few gigabytes of data, which further exacerbates this issue \cite{puech2007dicomworks}.

The most common approach for reducing access times is based on the combination of local cache and prefetching mechanisms that attempt to anticipate the user requests. Nevertheless, their effectiveness depends on accurately predicting what data will be requested. Traditional approaches for cache and prefetching in the medical imaging scenario are based on static rules over specific parameters \cite{huang2011pacs,bui2001problem,meyer1994picture}. However, this strategy has innumerous problems and limitations, as discussed in section \ref{sec:cachePref}.

This paper proposes an intelligent Cloud storage gateway for medical imaging repositories, focused on the reduction of communication latency. The proposed architecture is an improvement of a previous approach, having a combination of static rules with pattern recognition algorithms, enabling the system to adapt to user’s routines and behaviors, and is fully compliant with the Digital Imaging and Communications in Medicine (DICOM) standard. 
%
%

The remainder of this paper is structured as follows: section \ref{sec:section2} provides a brief overview of two main concepts (DICOM and PACS); section \ref{sec:cachePref} elaborates on the concepts of cache and prefetching, describing some approaches for both; in section \ref{sec:PropArc} the proposed architecture for the implemented system is explained; the experimental procedure followed for evaluating our proposal is detailed in section \ref{sec:ExpProcedure}, and the results obtained are illustrated in section \ref{sec:results}. Finally, section \ref{sec:conclusions} contains the concluding remarks regarding the developed work.

\subsection{Medical Imaging Laboratories}
\label{sec:section2}

Medical imaging processes are managed by systems called PACS. This kind of systems appeared in the early 1980’s as small systems composed mainly of an acquisition device, a visualization workstation, a small repository and a printer, having subsequently evolved to handle all digital medical imaging data produced in a healthcare institution.

\begin{figure}[t!]
 \centering
 \includegraphics[width=0.97\linewidth]{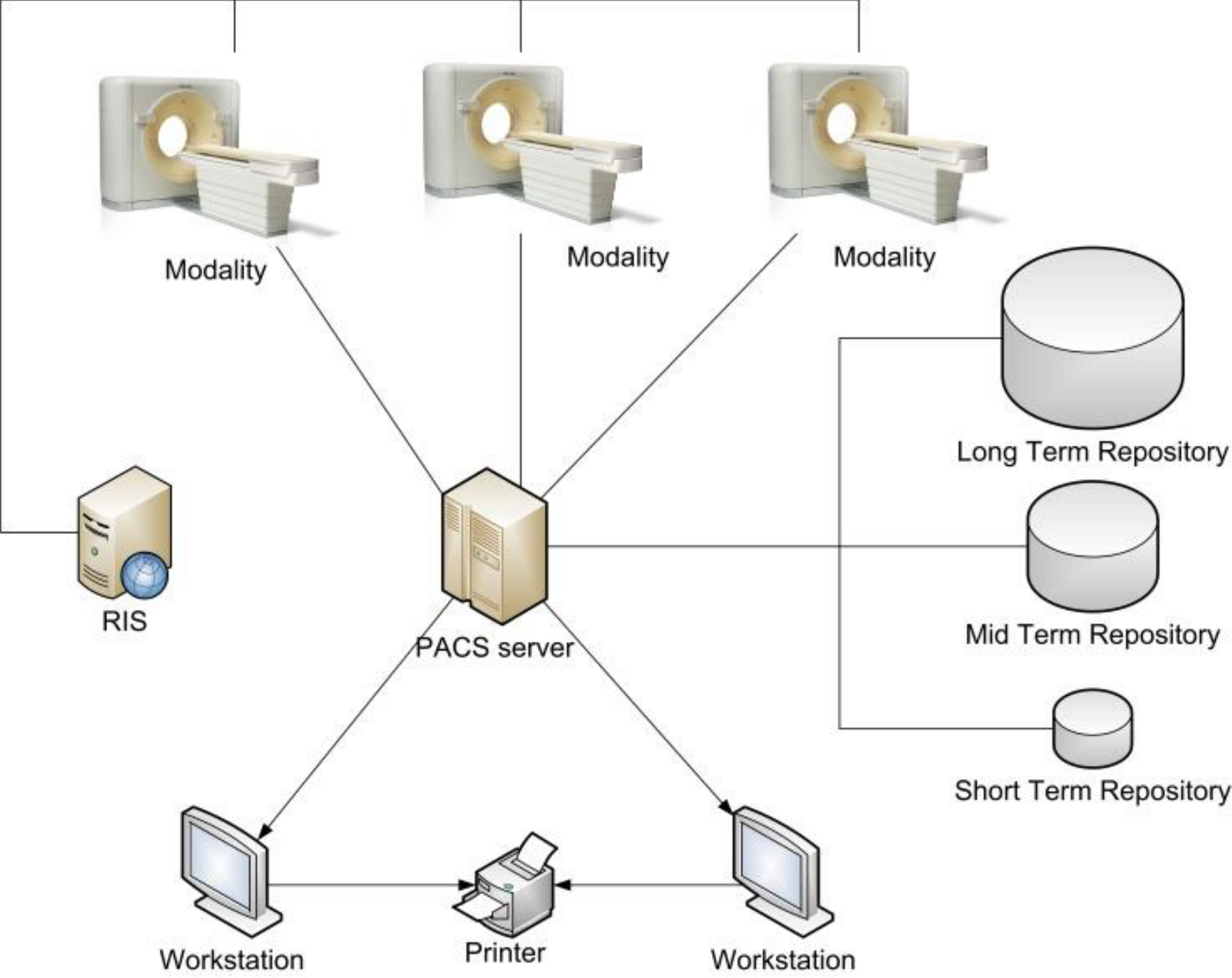}
 \caption{Typical PACS instance architecture.}
 \label{fig:PACS1}
\end{figure}

Figure \ref{fig:PACS1} shows a typical PACS instance that includes acquisition devices (i.e. the modalities), the repositories, PACS server, visualization workstations, printer, and the Radiology Information System (RIS).\\

\subsubsection{Digital Imaging and Communications in Medicine (DICOM)}

In the eighties, PACS were built in an ad-hoc fashion with proprietary communication protocols and file formats, among other aspects. For that reason, systems from distinct manufacturers, and in some cases even from the same manufacturer, were not interoperable, hindering the aggregation of all institutional devices into a single system capable of handling all medical imaging data. An international normalization effort to address these limitations resulted in the Digital Imaging and Communications in Medicine (DICOM) standard \cite{8}. DICOM defines data structure formats and communication processes, and introduces the DICOM Information Model (DIM) \cite{9} that outlines how relationships between real-world objects, such as studies and patients, must be represented. Moreover, it defines a set of network service commands for storing (C-Store), requesting (C-Get), querying (C-Find) and moving (C-Move) DICOM objects \cite{10}.

The DICOM network nodes are identified by their application entity title (AETitle) \cite{11} and the communications are done as a three-part process: association negotiation, service request/response and, in the end, the association release.

Because of its data encoding flexibility and the wide range of processes supported, DICOM was very well accepted in the medical imaging field. Nowadays, practically all devices follow this standard.\\

\subsubsection{PACS Outsourcing}
\label{subsec:PACSoutsource}

\begin{figure}[h!]
 \centering
 \includegraphics[width=0.93\linewidth]{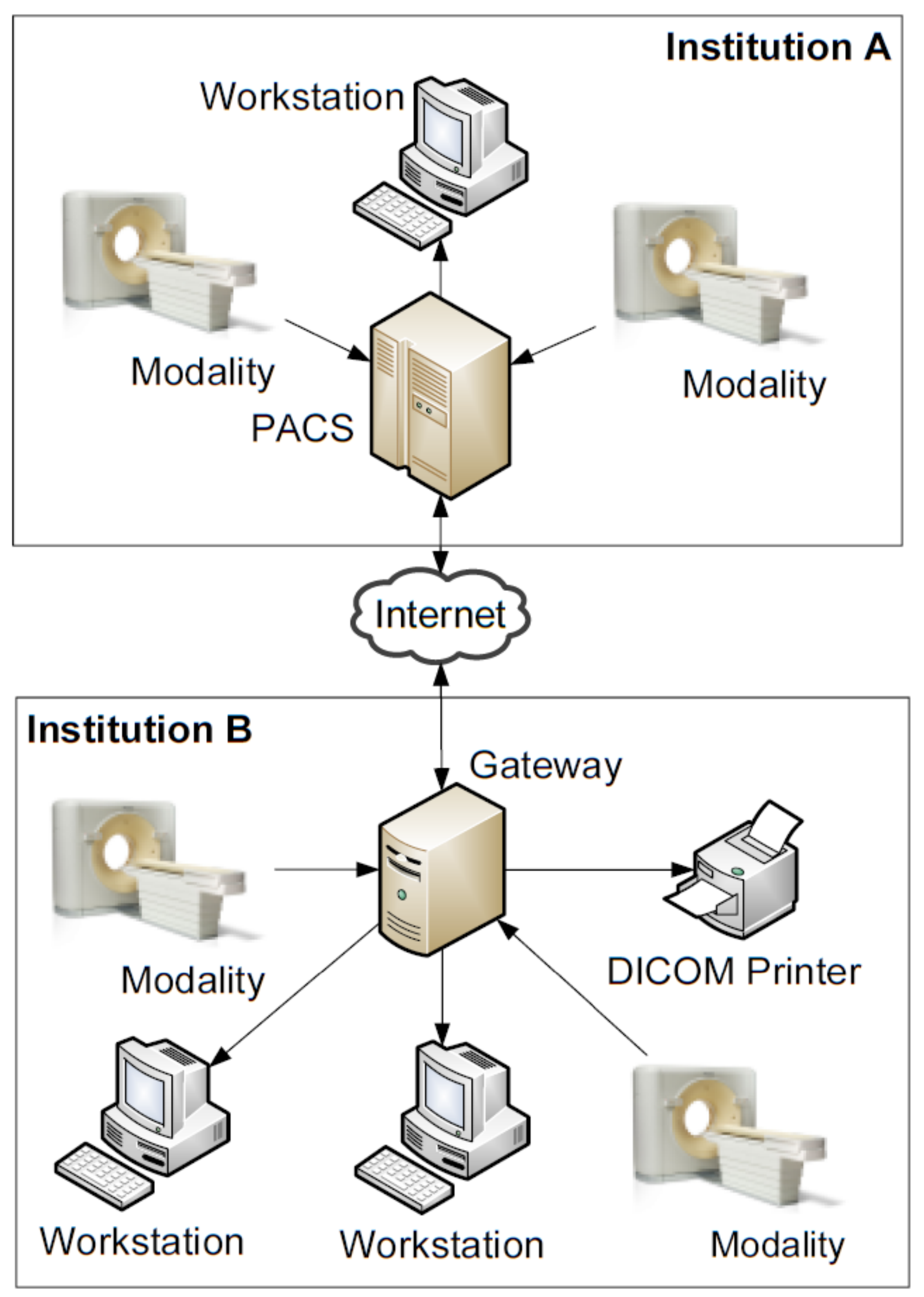}
 \caption{Architecture of the system deployed in two Portuguese healthcare institutions, described in \cite{silva2013agile}.}
 \label{fig:PACS2}
\end{figure}

Usually, PACS are constrained to a single institution. Nevertheless, the Cloud and the proliferation of high-speed Internet connections created the means to broaden PACS horizons. For instance, it is now possible to deploy a Regional PACS over the Cloud or federated distributed facilities.
In a previous work \cite{silva2013agile}, we described a federating system  for two clinics (institution A and B in Figure \ref{fig:PACS2}). In this setting, the central PACS archive is hosted on a private Cloud located at institution A while institution B only has a gateway that communicates with the PACS server via Internet.

The main concern with the solution depicted in Figure \ref{fig:PACS2} is that although some Internet connections can already provide an acceptable quality of service, these cannot compete with an Intranet based solution in terms of bandwidth and data transfer speeds. This constraint is hindering the adoption of PACS Cloud solutions by the institutions \cite{philbin2011will}. This issue could be minimized by endowing the gateway of institution B with a cache\cite{silva2014centralized}, but nevertheless the question of how to populate and evict the cache in an effective way still remained.

\subsection{Cache and Prefetching}
\label{sec:cachePref}

Cache effectiveness depends on several factors, namely: (1) the probability of finding the needed data in the cache; (2) length of time needed to retrieve data from the cache; (3) delay introduced by the cache processes when it does not have the requested data and (4) overheads due to maintenance of cache consistency \cite{smith1982cache}.
In this work we focused on the first aspect, namely on maximizing the likelihood of finding the needed data in the cache, trying to achieve better results than a previous architecture . Apart from the size of the cache, the two main contributing factors are the strategies for populating the cache and for selecting objects to be discarded when the cache is full. 
%
%
Cache population can be achieved by a passive mode, in which objects are stored in cache when they are first fetched from the source, or by prefetching, which consists on predicting future requests and retrieving the corresponding data before the requests actually occur. Passive cache population is most commonly used, and works well in scenarios that have high probability of repeating requests for a same object \cite{bui2001problem}.
The cache eviction process can follow numerous approaches known as cache replacement policies, among which Least Recently Used (LRU) and Least Frequently Used (LFU) are common options. 

Some PACS archives are hierarchically organized in short-term, mid-term and long-term repositories (Figure \ref{fig:PACS1}). The long-term repository uses cheap and slow technologies to store all studies, while the mid-term and short-term repositories provide faster access times but only keep copies of some of the objects stored in the long-term repository \cite{huang2011pacs}. These partial repositories are usually populated according to static rules associated with patient appointments, patient’s age, examination type, amongst other characteristics \cite{huang2011pacs,bui2001problem}.

The main limitation of strategies based on static rules is that they have to be tailored for each institution, taking into account the workflow, software and user’s behavior. Furthermore, such rules have to accommodate all possible situations, which can lead to populating the cache with unnecessary data. For instance, Bui et al. describe a prefetching mechanism with 100\% recall but only 50\% precision \cite{bui2001problem}, which means that all data is in cache when needed but only 50\% of the prefetched data is actually needed. Static rules have also the potential of producing perverse results in some situations. In a commercial solution, for example, we observed that a prefetching mechanism of a cache gateway associated to an outsourced PACS archive prefetches all results for all queries performed. So, the cache is populated with undesired studies when a user makes a bad query. Another reported case was related to the “poisoning” of cache population with massive number of studies requested by a user performing a non-standard task. For instance, an auditor requesting all CT (computed tomography) studies performed in the previous year could fill the cache with undesired studies, forcing all other requests to be served directly from the remote archive with consequent delay. Besides these considerations, traditional prefetching mechanisms can also overload the remote repository with requests and be stressing to the network.

Pattern recognition and machine learning have been increasingly used for cache replacement policies as well as prefetching. For instance, Pal and Jain \cite{pal2014approach} proposed a prefetching mechanism for web browsing that uses Markov models to predict which pages the users will request next. On another reported case, Garcia et al. \cite{garcia2013neural} used neural networks to predict the part of the map would be needed next, for map and navigation services.
Hybrid approaches for cache replacement and prefetching have also been described, namely by combining traditional approaches, such as LRU and LFU, with machine learning algorithms. An example is the application of machine learning to enhance conventional cache replacement policies for web browsing, as described by Ali et al. \cite{ali2012intelligent}. However, reports about the exploration of hybrid approaches in medical imaging environments are not found in the literature.

\section{Material and methods}
\subsection{Proposed Architecture}
\label{sec:PropArc}

This section proposes and describes an architecture of a new Cloud storage gateway (Figure 3) that supports cache replacement and prefetching for distributed medical imaging environments, aimed at minimizing the communication latency by learning the behavior of users. The solution was built to work with the gateway showed in the Figure 2.

\begin{figure}[h!]
 \centering
 \includegraphics[width=0.95\linewidth]{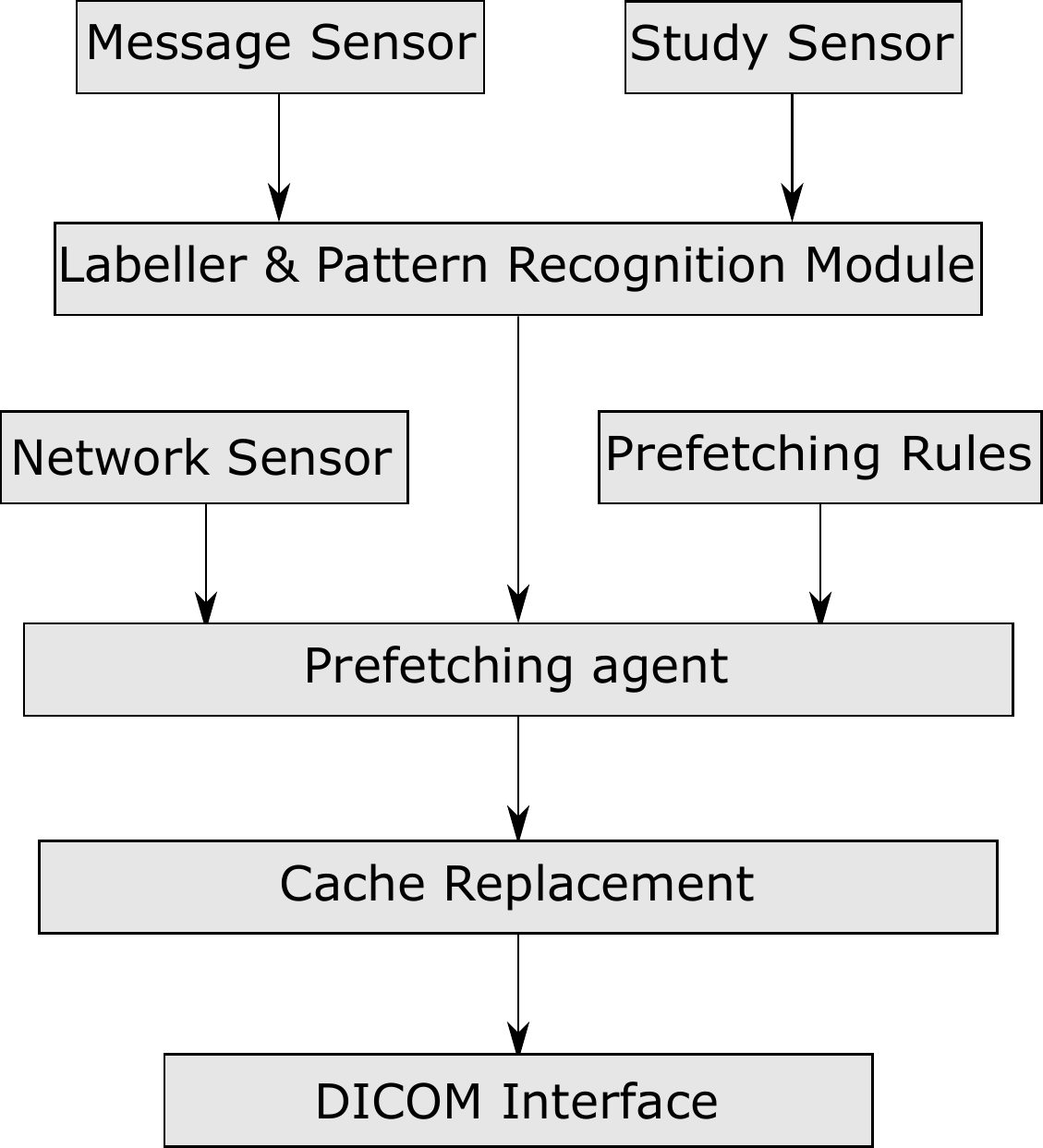}
 \caption{Schematic diagram of the proposed architecture.}
 \label{fig:ProposedArch}
\end{figure}

\subsubsection{Sensors}

The proposed system performs predictions based on environmental conditions, and so it must be equipped with distinct types of sensors to capture those conditions.

%
%

\subsubsection*{Message Sensor}

This sensor is the most important source for the new pattern recognition system because it allows capturing the messages interchanged between the local area network and the remote cloud archive. Since the proposed system has direct access to the network messages interchanged through the gateway, a set of listeners were integrated in the gateway, being awakened every time a new message is sent to (or received from) the repository.

The listeners generate a event report every time a message is exchanged. A set of metadata is saved in a log file, including the following information elements:

\begin{itemize}
\item time and date of message; 
\item kind of request;
\item UID of the study requested or the query made;
\item requesting application entity;
\item destination application entity.
\end{itemize}

The response messages contain additional information elements that are also registered in this log, such as the identifiers of the studies that match a query. All this information is then used in the pattern recognition system, as explained in \ref{subsec:labeller}.
%
%

\subsubsection*{Study Sensor}
In some cases, the network data exchanged is not enough to assess the usage pattern. For that reason, a study sensor that can query the repository in order to extract characteristics regarding a specific study was also implemented.

\subsubsection*{Network Sensor}
\label{subsec:NetSen}

In order to optimize our system, a new sensor was developed from scratch, to allow assessing the network conditions before proceeding with prefetching. This component monitors the Cloud gateway communications, watching the network requests and responses. This way, this sensor is able to continuously assess if the network is overloaded or if it can support additional traffic generated by the prefetching process. If the network is stressed, prefetching usage could have a harmful effect on the performance of the system. This new sensor is essential for a correct deployment of the new mechanisms that are explored in this article, allowing a more robust and adequate performance of cache replacement.\\
%
%

\subsubsection{Labeller \& Pattern Recognition Module}
\label{subsec:labeller}

This module is responsible for detecting which usage pattern best fits the user’s behavior and results from the integration of a new Labeller module within an adapted version of the Pattern Recognition mechanism proposed in \cite{Carlos}, classifying user behavior to understand if it is relevant to prefetching mechanisms or an one-off event with little medical and health relevance.
%
%

The key idea is to classify the behavior of the healthcare professionals based on the number and relationships of requested studies following each DICOM C-Find (i.e. query). To achieve this, healthcare professionals’ interactions were categorized into four distinct usage behavior patterns:

\begin{itemize}
\item Patient revising (class 1):  user is revising the studies of a single patient. This is potentially a usage pattern that demands fast access to images to preserve the quality of service;
\item Modality revising (class 2): user is revising the studies of a specific modality. This scenario is more critical for "heavier" modalities; 
\item Inconsequent query (class 3): pattern representative of user error situations or queries that do not result in download of imaging data.
\item “Other” usage (class 4): pattern representative of usage scenarios not identified by this architecture, such as an auditor evaluating all images of a whole department in a certain time window.
\end{itemize}

The functioning of this "module" proceeds as follows:  firstly, it splits the events sensed by the Message Sensor by AETitle (i.e. host). After that, it pre-processes and performs feature extraction on that information. The result of this process is used in two distinct ways: (1) it is sent to a set of trained MultiLayer Perceptron (MLP) models so they can predict which usage pattern best fits the user’s behavior and (2) it is saved to a log file for training the models. Based on the evaluation performed in a previous work, the models are trained by incremental learning, leading to more representive and real-based results. For this, at the end of each day, each training instance (query event) is assigned to the corresponding usage pattern based on the consequent study requests.

The features used by the models are of three kinds: (1) time features, e.g. the hour of the day, the day of the month, the month; (2) history features, e.g. number of patterns of each kind previous to this pattern, the last pattern, time since the last pattern; and (3) type of query, i.e. the parameters embedded in the query.\\

\subsubsection{Cache Replacement}

This module has two distinct agents: a cache manager agent, responsible for managing the imaging data stored in the cache system, and an eviction agent, responsible for dumping objects that are not necessary when the cache is full or has reached an occupation ratio that could hinder the storage of new objects. This eviction agent uses a LRU function, a broadly used solution in cache management which is relatively simple to implement and provides good performance \cite{LRU}, hence being the golden standard and usually used in real-environment. The LRU function assigns a weight of 100 to the newest study in cache and 0 to the oldest one. The remaining studies are assigned a value that corresponds to the ratio between its distance to the oldest study and its distance to the newest study.
%
%

The cache manager agent is important since it connects the repository and the database. The repository stores the image data as blobs of information, which are not processed and not searchable. A relational database is used to store information related to the repository objects. This way, it is possible to assess the amount of data stored, which images are stored, from what studies, from what patients, and also the amount of time that the images have been stored in the cache, allowing a better behavior of the eviction agent.\\

\subsubsection{Prefetching agent}

The prefetching agent has two distinct levels: short-term prefetching and long-term prefetching.

The long-term mode is triggered using the information from Network Sensor. When the system has low usage level, for instance during nights and weekends, this type of prefetching is used. Nonetheless, each time a user requests a study, this module extracts some characteristics of the study, such as the modality and production date (last day, week, month or year). Every characteristic has a counter associated with, and when the long-term prefetching is triggered, it uses the counters to infer which subset of images will more likely be requested, for instance: CTs of the last two months. As such, if the cache has free-space and the network conditions allow, the prefetching agent requests all studies that match the most popular categories of images.

The short-term mode uses the time between the query and the requests for prefetching the studies before being solicited by the user. The pattern recognition agent is used in this process. It is triggered when a query is made and predicts which usage pattern best fits the user’s behavior. After that, two parallel processes are executed:

\begin{enumerate}
\item The results of the user query are evaluated and the ones that match the usage pattern are selected for prefetching. Prefetching rules are applied in order to assess which ones have higher priority.
\item The prefetching agent instantly makes a query to the repository, based on the new outputs of the pattern recognition module. The query depends on the predicted usage pattern: if this is “patient revising” the query is performed by patient ID; if the usage pattern is “modality revising”, a query by modality with a time window of one month is performed. The two remaining usage patterns are not considered for prefetching. After that, all query results are evaluated by the prefetching rules and the ones with higher score are prefetched, if not already covered by process 1.
\end{enumerate}
This agent is essential to improve the system and have a more adequate and robust solution for the problem in question, as further demonstrated in results, when compared with the golden standard LRU.  
%
%
\subsubsection*{Prefetching Rule}

\label{subsec::PrefetchRule}

In order for the prefetching agent to know which studies should be fetched, a neural network function that learns with the AETitle history is provided. It uses a MLP neural network for each DICOM node, with the following inputs: the length of time since the study was produced, the body part, the modality, the patient gender, the patient age, the usage pattern and the institution that procuded the examination.

The neural networks are trained each day (or week) using as training data the studies retrieved by the searches. These are labeled as positive instances if requested after the search, and as negative instances otherwise. The output of the function is a measure of the likelihood of that study being requested.\\

\subsubsection{DICOM Interface}

As previously described, some modules of this architecture need to communicate with third-party PACS equipment. For that reason, the system provides the DICOM interface module. This middleware converts module requests into DICOM requests and sends them to the destination.

\subsection{Experimental Evaluation}
\label{sec:ExpProcedure}

Testing the proposed architecture in distinct healthcare institutions under different environment conditions (network, user schedules, number of workstations and so on) could be dangerous for the regular institutional processes, since some of the tests would overload the servers and network.
For this reason we performed our tests through simulations, under different conditions, over a real-world dataset. Each distinct scenario was simulated ten times to minimize the impact of random initialization of some parameters of the system, such as the MLPs. For each condition, two metrics were used to analyze system performance:

\begin{itemize}
\item Hit Ratio, calculated by dividing the number of times a requested object was stored in the cache by the number of object requests.
\item Retrieval time per image, calculated by dividing the total time needed to retrieve a requested study by the number of object requests. \\
\end{itemize}

\subsubsection{Real-World Dataset}

The real-world dataset is composed of two parts: (1) an XML file containing information about the messages exchanged in the network and (2) an index containing anonymized information about the studies stored in the institutions.

The XML file contains data for 5186 DICOM requests: C-Move and C-Find. These requests are from the workstations to the PACS server, during a 3-month period. Nevertheless, the simulation needed more information, including: (1) the size of the studies; (2) the number of files; (3) the results retrieved for a query; (4) the anonymized patient data; and (5) the characteristics of the study.
For that, we used a replica of the repository database, with sensitive data removed by applying a one-way function (i.e. hash) to some of the fields, for instance: patient name. The same function was also applied to the XML file, in order to hide some parameters of the queries that contain private data. This strategy allowed us to reproduce the results retrieved by a user’s query, without having access to the actual raw data.

This dataset was extracted from the Cloud gateway system described in section \ref{subsec:PACSoutsource}, which has three remote workstations on Institution B. The shared repository holds 2 terabytes of medical imaging studies. For the validation experiments, the static rules already used in this real-world gateway were  imported and applied in our system.\\

\subsubsection{Test conditions}

Tests were executed simulating distinct situations, including cache size. For that, and considering the size of the dataset, cache sizes of 2.5, 10, 20, 50 and 100 gigabytes were used. For each cache size, the system was tested with the following configurations:

\begin{itemize}
\item Configuration 1: LRU is selected as eviction policy, and no prefetching policy is used, as explained in \ref{subsec:RelatedWork}, representing the golden standard in the area.	
%
%
\item Configuration 2: LRU is selected as eviction policy, and both short and long-term prefetching modes are used. 
\end{itemize}

Additionally, all configurations have a passive cache population mode, which means that all studies that pass through the gateway are also stored in the cache.

\subsection{Search for similar or related works}
%
%
\label{subsec:RelatedWork}

Since this a area of growing interest, an exhaustive search of material and related works was performed. Although several works are published in the area of cloud storage in health/medical environment, none of the literature in the forums (Google Scholar, Web of Science, IEEE Xplore Digital Library) compared the hit ratio with a growing cache size and retrieval time with a growing cache size, resourcing in other measurements that, in our opinion, are less relevant to this area and the objective of this work. So LRU was selected to compare to our system, since it is considered the golden standard in the area. Furthermore, for a more adequate and fair comparison, LRU was performed with the pipeline of our system, with modifications to improve its results. 

\begin{figure*}[h]
    \centering
       \includegraphics[width=0.79\textwidth]{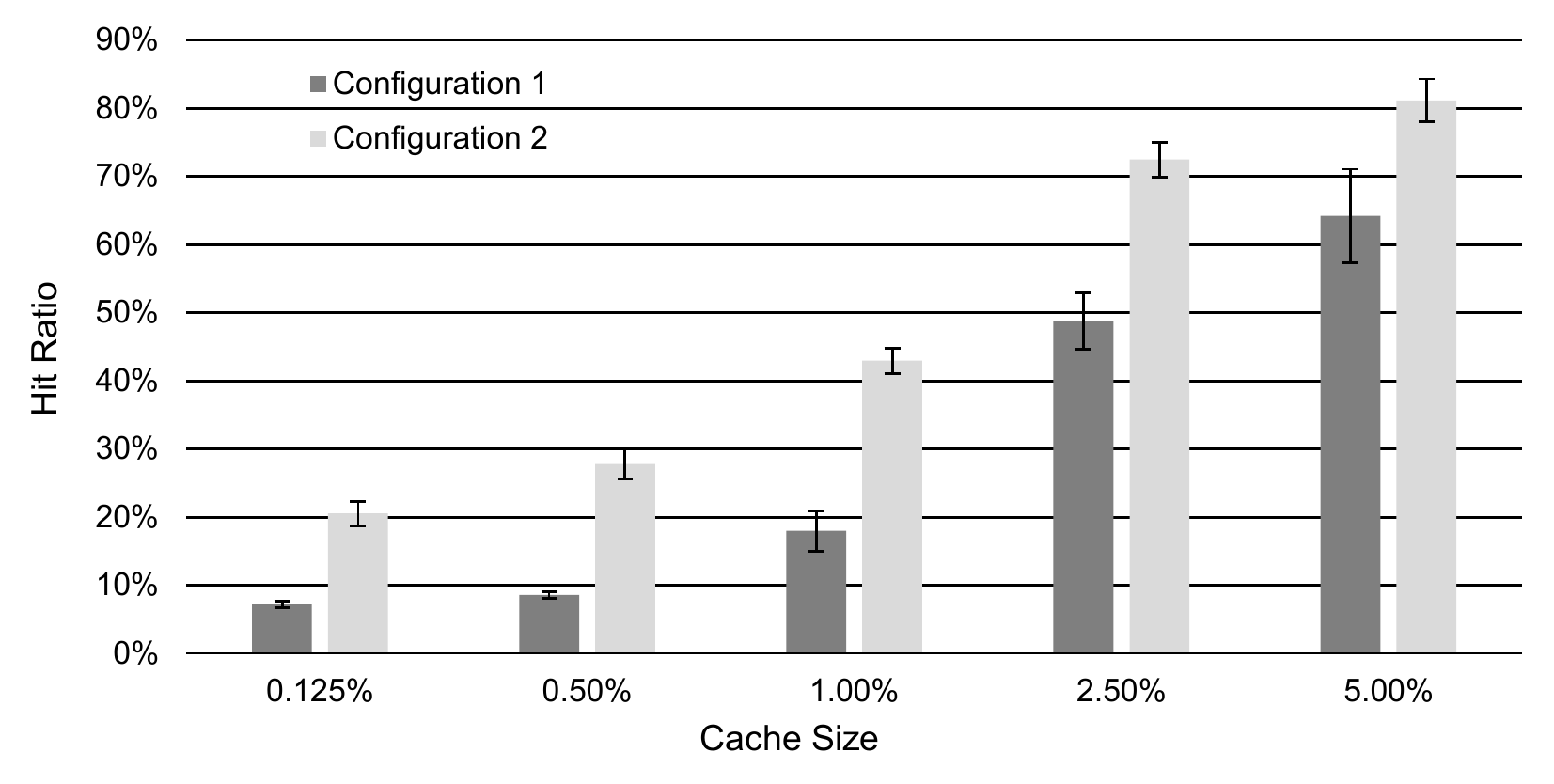}
        \caption{Hit Ratios, in percentage, for both configurations and for different cache sizes. Cache size is presented as a fraction of the total dataset.}
        \label{fig:GraficoHitRatio}
\end{figure*}

\begin{figure*}[h]
   \centering
        \includegraphics[width=0.79\textwidth]{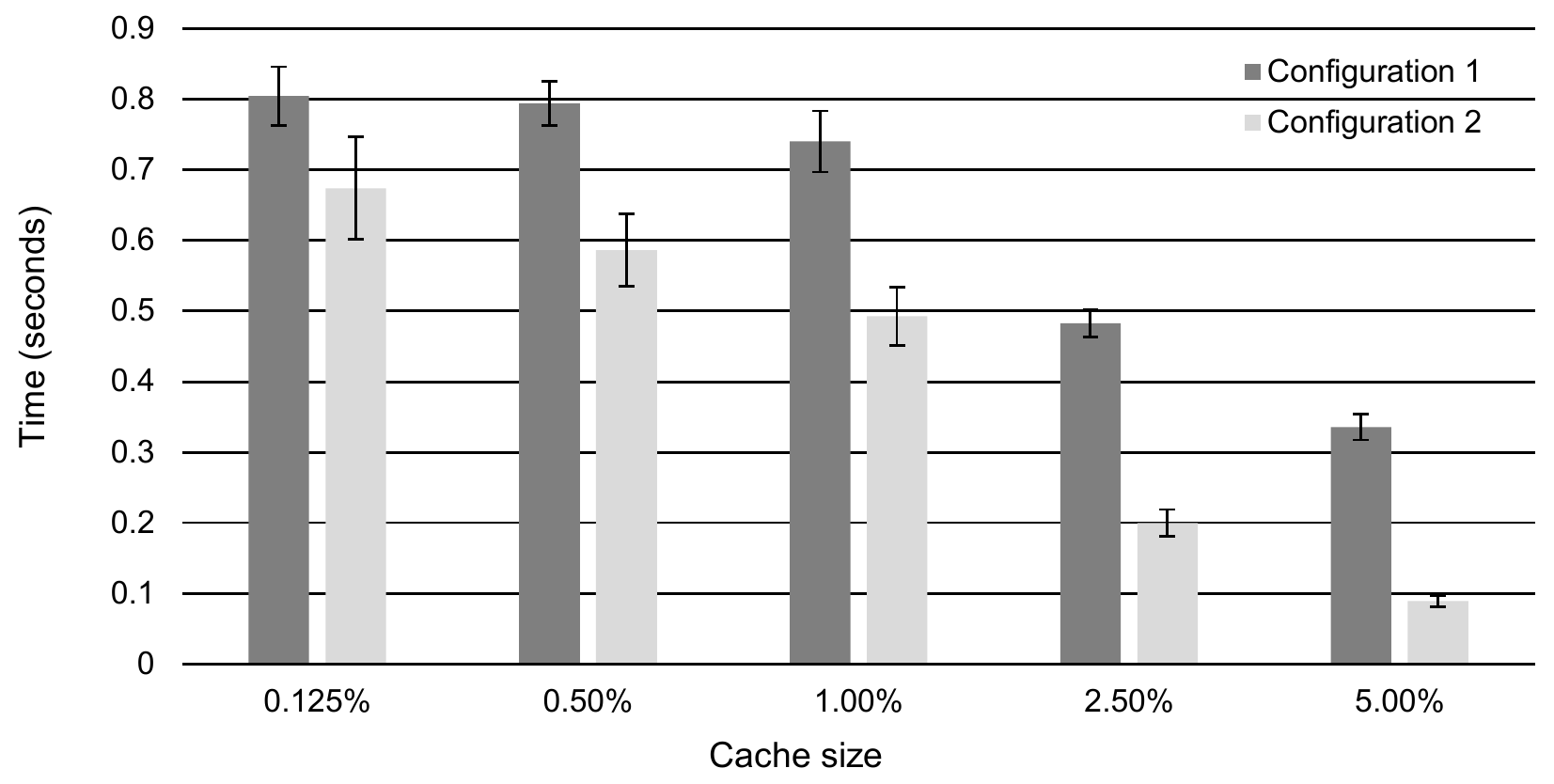}
        \caption{Retrieval Times, in seconds, for both configurations and for different cache sizes. Cache size is presented as a fraction of the total dataset.}
        \label{fig:GraficoTempo}
\end{figure*}

\section{Results \& Discussion}
\label{sec:results}

Numerous tests were carried out to evaluate the effectiveness of the proposal. Configuration 1 was used as the reference for comparison with the proposed system, since LRU is a widely used eviction approach which provides good results with relatively simple implementation \cite{LRU} and as been the golden standard in this area. In addition to LRU, our system uses also short and long-term prefetching modes, needing for that a Network Sensor, as previously explained. This configuration have lower computational cost when compared with more complex configurations, allowing its application in different systems with low impact in their performance.



While there is no rule of thumb for defining the necessary cache size for a hospital information system, and literature supporting appropriate cache sizing procedure is scarce, some literature recommends that such storage component should cover actual imaging of 1 month up to 3 months \cite{Beatiful}, which can reach several hundred gigabytes of data \cite{beutel2000handbook2}. Despite being desirable to have a large cache, this brings significant costs to the system \cite{beutel2000handbook2}, thus, adequate cache size should be chosen so that its size is "optimized" regarding the specific workflow of the system where it will be implemented (e.g. in a RIS) \cite{Beatiful}.  So, we opted to analyse the performance of the proposed system with a range of smaller cache sizes, which go up to a maximum of 100 gigabytes. 

The results obtained for hit ratio and retrieval time per image are shown in Figure \ref{fig:GraficoHitRatio} and Figure \ref{fig:GraficoTempo}, respectively. Cache sizes are presented as percentages, representing the corresponding fraction of the dataset instead of the actual value in gigabytes.

Analysing hit ratio, it is possible to observe in Figure \ref{fig:GraficoHitRatio} that the proposed system exhibits higher hit ratio than the "baseline" configuration for every cache size tested, with the maximum value being a hit ratio of approximately 81\% for a cache size of 5\% (equivalent to 100 gigabytes). In what concerns retrieval time per image, time reduces with the increase in cache size, as expected. Moreover, the proposed system has lower retrieval times when compared to the base system, for every cache size considered in this work.

These results show that the proposed implementation improves system performance in every tested scenario, when compared to configuration 1. In fact, with a cache size of 100 gigabytes, the proposed system required on average 73\% less time to retrieve each image, when compared to configuration 1. This shows that using short and long-term prefetching modes can considerable decrease the impact of communication latency. Furthermore, a hit ratio above 80\% is very significant taking into account that 100 gigabytes represent only 5\% of the total data stored in the main repository that is remotely located.

Moreover, when comparing both configurations in the larger cache spectrum, we can see that our system achieves a higher hit ratio and 33\% lower retrieval time using a cache size of 2.5\%, than the baseline configuration with a cache with double the size. These results show that our system proposal brings considerable performance benefits even with a smaller cache.

  Another interesting aspect to take into account is the fact that static rules, which were imported from the real-world validation scenario, apply specifically to the more common cases (classes 1 \& 2, described in \ref{subsec:labeller}) that cover, in the dataset used, approximately 80\% of the user requests. While these static rules allow a more efficient system response in typical requests, the remaining 20\% are not contemplated. Due to the adaptive nature of our architecture, distinct behavior patterns are correctly detected and classified, enabling the system not only to match and even improve the performance for typical requests, but also to greatly improve performance for the remaining requests, which would normally be processed with a basic configuration similar to that of configuration 1.

It is important to refer that this work focused mainly on the small-cache regime, where hit ratios vary more significantly depending on the caching policies that are used. The small-cache regime is of particular interest, since it is a regime where the use of eviction policies coupled with other policies, namely prefetching policies, produces significantly better performances compared to the sole use of eviction policies \cite{LRU}. As aforementioned, this was effectively observed, as configuration 2 outperformed configuration 1 in every tested scenario.

\section{Conclusions}
\label{sec:conclusions}

In this paper, we propose an intelligent Cloud storage gateway that supports prefetching and eviction policies, aiming to reduce the communication latency when accessing remote medical imaging repositories. This scenario is particularly important due to the current trend for outsourcing PACS archives to the Cloud.

The proposed architecture uses a set of prefetching and eviction policies. In what concerns eviction, LRU was the selected policy. Regarding prefetching, two prefetching modes were used: long-term and short-term prefetching. The first is responsible for fetching objects that will probably be requested in the next day or week, whereas the latter is for more immediate needs, i.e. for the next minutes.

The system was subjected to exhaustive tests over a real-world dataset, leading to observed reduction of image retrieval times close to, or even over, 60\% for the larger cache sizes. The results obtained show that the combined use of eviction and prefetching policies proposed in this paper can significantly reduce communication latency, even with a considerably reduced cache in comparison to the total size of the main repository (small-cache regime).

The hybrid solution herein proposed yields a system capable of adjusting to the distinct, specific workflows of different institutions, whilst offering significant improvements in system performance, namely regarding hit ratio and retrieval time metrics.

\section{Compliance with Ethical Standards}

\textbf{Funding} 

This work has received support from the  ERDF – European Regional Development Fund through the Operational Programme for Competitiveness and Internationalisation - COMPETE 2020 Programme, and by National Funds through the FCT – Fundação para a Ciência e a Tecnologia within project PTDC/EEI-ESS/6815/2014; POCI-01-0145-FEDER-016694. Sérgio Matos is funded under the FCT Investigator programme.

\textbf{Conflict of Interest}

All authors declare that there are no conflicts of interest in this work.

\textbf{Ethical Approval}

This article does not contain any studies with human participants or animals performed by any of the authors.

\bibliographystyle{ieeetr}
\bibliography{ref}
\end{document}